\documentclass[journal]{vgtc}                
\ifpdf
  \pdfoutput=1\relax                   
  \usepackage{graphicx}                
  \DeclareGraphicsExtensions{.pdf,.png,.jpg,.jpeg} 
\else
  \usepackage{graphicx}                
  \DeclareGraphicsExtensions{.eps}     
\fi%

\graphicspath{{figures/}{pictures/}{images/}{./}} 

\usepackage{microtype}                 
\PassOptionsToPackage{warn}{textcomp}  
\usepackage{textcomp}                  
\usepackage{mathptmx}                  
\usepackage{times}                     
\usepackage{cite}                      
\usepackage{tabu}                      
\usepackage{booktabs}                  

\usepackage[frozencache,cachedir=.]{minted}
\usepackage{enumitem}
\setlist{noitemsep} 

\newcommand{\sys}[1]{\textit{Quick Dashboard}}
\MakeRobust{\sys}
\setminted{fontsize=\small,baselinestretch=1}

\newcommand{\edit}[1]{{#1}}                  

\onlineid{1017}

\vgtccategory{Research}
\vgtcpapertype{system}

\title{A Declarative Specification for Authoring Metrics Dashboards}

\author{Will Epperson, Kanit Wongsuphasawat, Allison Whilden, Fan Du, Justin Talbot}
\authorfooter{
\item Will Epperson is with Carnegie Mellon University. E-mail: willepp@cmu.edu.
\item Kanit Wongsuphasawat is with Databricks. E-mail: kanit.w@databricks.com.
\item Allison Whilden is with Databricks. E-mail: allison.whilden@databricks.com.
\item Fan Du is with Databricks. E-mail: fan.du@databricks.com.
\item Justin Talbot is with Databricks. E-mail: justin.talbot@databricks.com.
}

\shortauthortitle{Epperson \MakeLowercase{\textit{et al.}}: A Declarative Specification for Authoring Metrics Dashboards}

\abstract{
Despite their ubiquity, authoring dashboards for metrics reporting in modern data analysis tools remains a manual, time-consuming process.
Rather than focusing on interesting combinations of their data, users have to spend time creating each chart in a dashboard one by one.
This makes dashboard creation slow and tedious.
We conducted a review of production metrics dashboards and found that many dashboards contain a common structure: breaking down one or more metrics by different dimensions. 
In response, we developed a high-level specification for describing dashboards as sections of metrics repeated across the same dimensions and a graphical interface, \sys{}, for authoring dashboards based on this specification.
We present several usage examples that demonstrate the flexibility of this specification to create various kinds of dashboards and support a data-first approach to dashboard authoring.
}


\CCScatlist{ 
 \CCScat{K.6.1}{Management of Computing and Information Systems}%
{Project and People Management}{Life Cycle};
 \CCScat{K.7.m}{The Computing Profession}{Miscellaneous}{Ethics}
}

\teaser{
  \centering
  \includegraphics[width=\linewidth]{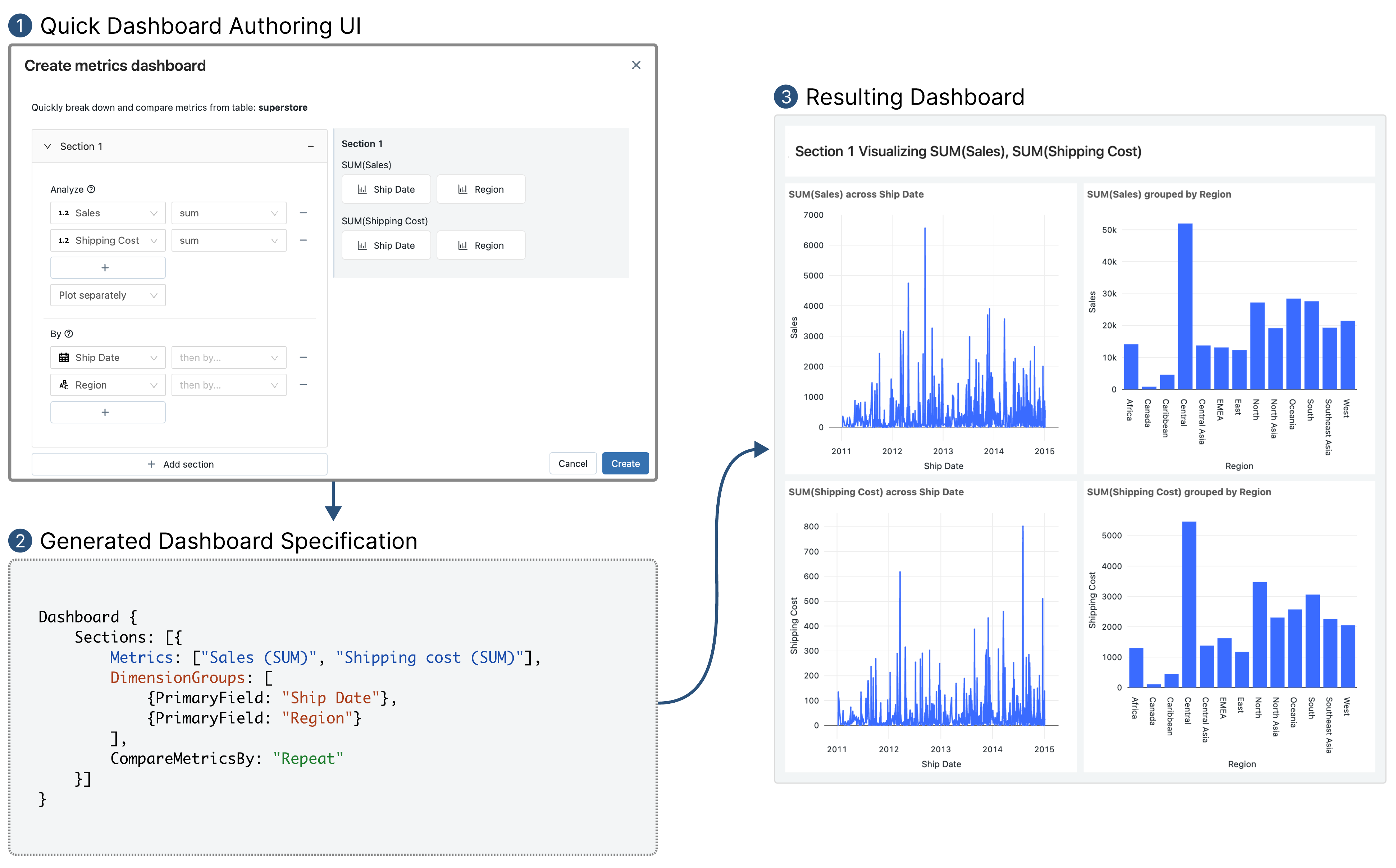}
  \caption{The \sys{} system allows users to rapidly specify dashboards by choosing metrics and dimensions of interest from a data table. \textbf{(1)} A user begins by using the \sys{} UI to choose metrics and dimensions for each section of their dashboard. In this example, the user has selected two metrics (Sales and Shipping Cost) and two dimensions (Ship Date and Region). \textbf{(2)} The UI creates a corresponding dashboard specification for the user's input. \textbf{(3)} The system then generates the final dashboard by combining each metric and dimension according to the specification. This dashboard can be used immediately or further customized by changing the chart types or re-arranging components.}
  \label{fig:teaser}
}

\begin{document}


\maketitle

\section{Introduction}



Dashboards are used for a variety of purposes, including monitoring important metrics, providing an overview of data, or for communication \cite{Sarikaya2019WhatDW}.
In this paper, we focus on using dashboards to monitor and report on metrics of interest.
Metrics are quantitative variables such as the total dollar amount of sales or number of successful surgeries.
When analyzed across \textit{temporal} or \textit{categorical} dimensions, interesting trends can be found.
For example, a dashboard might show metrics like the total number of sales or the ratio of successful to unsuccessful surgeries across dimensions like time and region.
Putting together multiple visualizations in a dashboard allows users to see how their metrics change across these different slices. 

Current dashboard authoring experiences such as Tableau, Looker, or PowerBI typically focus on the \textit{bottom-up} creation of individual charts, which are later combined into a dashboard \cite{tableau, googleLooker, powerBI}.
Whether creating dashboards through code-first or GUI interfaces, the bottom-up approach is slow and repetitive. 
Each chart must be manually created, saved, and then combined into the final dashboard. 
Even for a simple dashboard with only a few metrics and dimensions, the time spent on specifying the charts of a dashboard could be better spent on understanding the results and considering meaningful data combinations.
Recent research has discussed the steep learning curve for many dashboard tools that require users to have a deep knowledge of visualization design~\cite{Sarikaya2019WhatDW}.
Like other recent work~\cite{Pandey2022MEDLEYIR}, we view this as an opportunity for simpler interfaces to dashboard creation.

To simplify and accelerate dashboard creation, we observed the fact that dashboards are typically not a combination of random charts, but have an underlying repetitive structure.
Specifically, dashboards are often structured as sections of charts where the \textbf{same metrics are broken down by the same dimensions}.
We used this repetitive structure to design a high-level specification for describing dashboards.
\edit{To demonstrate the utility of this abstraction, we built a user interface, \sys{}, that allows users to quickly author dashboards.}
Our system allows users to specify \textit{which} parts of their data they are interested in visualizing rather than \textit{how} to visualize each combination of columns.
\sys{} uses a chart recommender to create each of the charts in the dashboard.
After creation, users can then customize each of the charts in the dashboard using a GUI chart editor and change the layout of the widgets.
\edit{Developers of dashboard authoring tools can use our specification to build experiences similar to that demonstrated in \sys{} that focus on data-first rather than chart-first authoring flows.}
In short, our paper makes the following contributions:

\begin{enumerate}
    \item A high-level specification for describing dashboards as sections of metrics combined with dimensions. Each section contains the cross-product of all metrics and dimensions.
    \item \edit{A demonstration of the flexibility and utility of the proposed dashboard specification through a no-code dashboard authoring tool, \sys{}, and examples of three real-world dashboards built using the tool.}
\end{enumerate}
\section{Related Work}

We draw from two primary areas of related work: visualization recommendations for single and multi-view charts and dedicated tools for dashboard authoring.

\subsection{Visualization Recommendation}
Visualization recommendation helps analysts understand their data by automating visual presentation or suggesting interesting parts of the data~\cite{2016-compassql}.
At the individual chart level, this aids in faster exploration of visual designs that best communicate the data \cite{mackinlayShowME2007}, and when looking at multiple charts can help analysts rapidly explore different aspects and combinations of data \cite{2016-voyager}.
\sys{} uses a rule-based chart recommender to generate individual charts that are composed into dashboards based on the data types of input fields, similar to \cite{compassQL2016}.
Since the focus of this paper is on a specification for expressing dashboards, we omit the fine-grained details on how the individual charts are generated however discuss input format for our chart recommender in \autoref{sec: chart recommendation}.
\sys{} can be used with other chart recommendation modules that take similar inputs.

Dashboards are instances of multi-view visualizations, where multiple charts are used with common fields to break down trends \cite{Roberts1998OnEM}.
Beginning with Tufte's early work on small multiples \cite{Tufte01}, recent research has explored guidelines and design principles for crafting effective multi-view visualizations \cite{Qu2018KeepingMV, Baldonado2000GuidelinesFU}.
We incorporate such guidelines into our system, in particular focusing on consistent scales and axes when multiple recommended charts have the same field in a section (constraint C1 from \cite{Qu2018KeepingMV}).
Future iterations of our system plan on further incorporating the guidance on using consistent colors across all charts in a dashboard from previous work \cite{Qu2018KeepingMV}.

\subsection{Dashboard Authoring Tools}
Another line of relevant research focuses on improved tools for dashboard creation. 
Recent approaches have explored supporting faster dashboard authoring through natural language interfaces~\cite{srinivasan2023bolt} or by providing example images to bootstrap dashboard creation~\cite{maLADV2021}.
Tools like VizDeck~\cite{Key2012VizDeckSD} focus on data \textit{exploration} by showing users many ranked charts that can be saved and combined into a dashboard.
Furthermore, multi-view recommendation approaches like MultiVision allow users to take a single visualization and augment it with other encodings and visualizations to create complementary views \cite{Wu2021MultiVisionDA}.
Our approach differs in that we assume the user has already explored their dataset and wants to focus on the presentation of specific columns compared to one another, namely metrics vs dimensions.
Therefore we focus on recommending an encoding of user-selected data as a dashboard, rather than supporting data discovery. 

Most relevant to our work is the Medley system that allows users to specify an analytic intent such as measure analysis and fields of interest and choose from recommended visualization groups \cite{Pandey2022MEDLEYIR}.
Our work differs in that we do not require users to specify their analysis goal upfront and we focus on a top-down specification of a dashboard. 
This makes the specification simpler since users do not have to select a task in addition to the fields they wish to visualize in their dashboard sections.
Additionally, the flexibility of our specification allows for the creation of dashboards with many of the same attribute combinations as Medley by describing sections of metrics and dimensions.
Future work could use our dashboard specification for task-based recommendations to support authoring experiences similar to Medley.
\section{Quick Dashboarding}


\begin{figure}
    \centering
    \includegraphics[height=.93\textheight, keepaspectratio ]{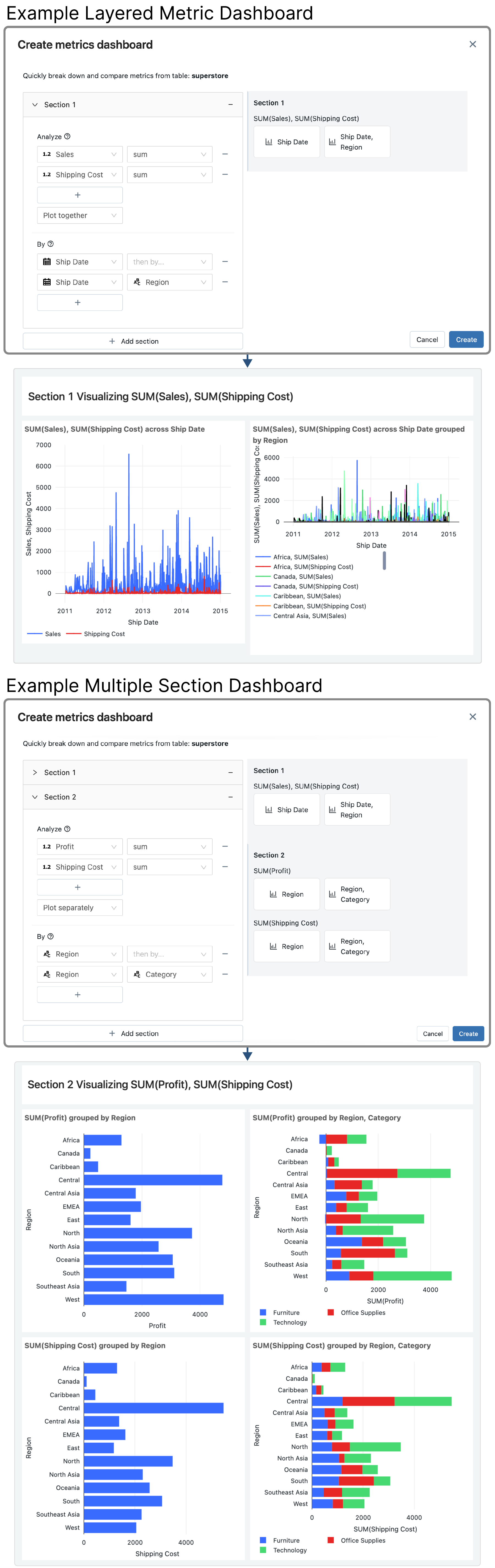}
    \caption{
    \textbf{Top:} A dashboard where two metrics, Sales and Shipping Cost, are layered together across two different dimension groups. 
    \textbf{Bottom:} A dashboard with multiple sections. The first section is the same as the top dashboard and thus not displayed in the preview.
    }
    \label{fig:example dashboards}
\end{figure}



To help users create dashboards by only specifying columns of interest from their data, we developed a novel dashboard specification and UI tool, \sys{}. 
In this section, we describe our process of developing the underlying dashboard specification and the UI interface. 
We also discuss several example dashboards that can be created with this specification.

\subsection{Goals \& Requirements}
In our discussions with current dashboard users at
a large technology company,
we found a common need was better support for quickly authoring a new dashboard without having to specify each component chart and query.
In order to support this data-driven and fast dashboard creation we identified the following requirements of our specification and system:

\begin{enumerate}
    \item [R1:] \textbf{Specify dashboard based on repeated metrics and dimensions:} Users should be able to create a dashboard by selecting which fields of a dataset they would like to present without having to specify \textit{how} to visualize each combination of fields.
    \item [R2:] \textbf{Allow customization afterward:} After we recommend an initial dashboard, users should be able to easily change chart types if they do not like the defaults.
    \item [R3:] \textbf{Single recommendation:} For each input specification, only a single dashboard should be generated. Since verifying dashboard designs is slow, we only show a single design to a user that they can edit rather than having them browse recommendations.
\end{enumerate}

\subsection{Developing the specification}
To generate a higher-level, data-focused, dashboard specification, we looked for repetitive structure among existing dashboards and considered the primary goals users have when creating dashboards.
We reviewed dozens of internal and customer dashboards and found that the dominant goal of dashboards was to report and break down \textit{metrics} of interest.
Metrics are business meaningful numbers such as the total number of sales.
To provide a more in-depth picture of metrics, dashboards often break them down by other aspects of the data.
For example, looking at the total number of sales across time or by region.
We refer to these fields used to break down metrics as dimensions.
Dashboards often have a logical structure to help make visual interpretation easier by grouping charts that break down the same metric into \textit{sections}.
These observed patterns informed our dashboard specification that we present in the next section.

\subsection{Dashboard Specification}
Our dashboard specification relies on two different types of fields (i.e. columns in a table): metrics and dimensions. 
This terminology is similar to many existing systems for data analysis and dashboard creation such as Tableau \cite{tableau}, although they refer to metrics as measures.
In our case, we use the following definitions:
\begin{itemize}
    \item \textit{Metric: } Metrics are quantitative fields that include their preferred aggregation (such as sum or mean).
    \item \textit{Dimension: } Dimensions are categorical or temporal fields that are used to break down metrics.
\end{itemize}
By selecting only metrics and dimensions of interest, users can fully specify their dashboard (R1).
We use the following specification to describe a dashboard:
\begin{minted}{js}
Dashboard {
    Sections: [{
        Metrics: Field[],
        DimensionGroups: [{PrimaryField, SecondaryField}],
        MetricLayout: "Layer" | "Repeat"
    }]
}
\end{minted}
There are several important details in this specification and how it is translated to an actual dashboard.
First, a dashboard is a series of sections, each with its own metrics and dimensions.
A dashboard can have an arbitrary number of sections, and each section has an arbitrary number of metrics and dimensions.
We require each section to have at least one metric.
Users can use \mintinline{js}{COUNT(*)} as a default metric to plot the count of dimensions if no other fields are specified as metrics.

In each section, all metrics are combined with all dimensions. 
The \mintinline{js}{MetricLayout} option determines \textit{how} metrics in the same section are combined. 
When this value is \mintinline{js}{"Layer"} then the metrics are all layered onto the same chart(s), which are then combined with the different dimension groups.
For example, \autoref{fig:example dashboards} (top) shows a dashboard where two metrics are combined in each chart.
When this value is \mintinline{js}{"Repeat"} then each metric is put in its own chart. 
\autoref{fig:teaser} and \autoref{fig:example dashboards} (bottom) show dashboards with each metric in its own chart.
By default, metrics are \textit{repeated} and plotted in separate charts.

Since each metric (or all metrics) are combined with each dimension, we use the concept of dimension \textit{groups}.
This allows us to create charts where a metric is broken down by multiple dimensions such as breaking down Sales by Date and Region in the same chart.
We limit the number of dimensions in a group to 2 since it makes encoding easier and putting more than 3 variables onto a single chart starts to become crowded and less informative.


\subsection{Chart Recommendation}
\label{sec: chart recommendation}

In order to produce charts from metrics and dimensions, we use a chart recommendation module to create charts that best visualize these fields.
Users do not have to choose the chart type or encoding channel, they simply specify their metrics and dimensions according to the specification above and we produce sensible charts for them (R1, R2).
We chose to recommend a single chart for each field input so that users do not have to browse options (R3), but can make modifications to the chart type and encoding afterward (R2).
Our chart recommendation module takes the following as inputs: each column and its datatype, along with a preferred axis.
This preferred axis allows us to force consistent axes for the same metric plotted across multiple dimensions.
The chart recommender has no notion of metric or dimension and uses a rule-based approach to choose effective visualizations based on the input column data types inspired by previous chart recommendation systems~\cite{2016-voyager, 2016-compassql}.

\subsection{\sys{} UI}
\edit{The \sys{} UI demonstrates how systems can leverage our specification to build dashboard authoring experiences.
\sys{} exists within a larger data analytics product, however similar authoring experiences can be developed in platforms that have access to data tables and can output dashboards.} 
\edit{In our authoring flow, users first select a table and then use the \sys{} authoring modal to select column combinations of interest. }
Therefore we show a dashboard preview on the right side of the interface so that users can understand how the metrics and dimensions they choose will be translated into a dashboard.
Once a user hits create, a dashboard specification is created which is then translated to an actual dashboard with the chart recommender.
In \autoref{fig:teaser} and \autoref{fig:example dashboards} we demonstrate a filled-in UI and the resulting dashboard for three different kinds of dashboards.
By using the UI, users do not have to manually author the dashboard specification.
This data-first approach to dashboard authoring also makes it easier for a broader set of users to create dashboards without the need to write queries to get their data or manually author each individual chart.

\subsection{Example Dashboards}
In this section, we describe several example dashboards and the specification that was used to create them.
These dashboard layouts are inspired by sections from real production dashboards at 
a large technology company
and demonstrate the flexibility of our specification in producing different kinds of sections.

\subsubsection{Example 1: Breaking down metrics by dimensions}
Our first example dashboard demonstrates our core abstraction of breaking down multiple metrics by multiple dimensions.
The UI, spec, and produced dashboard are shown in \autoref{fig:teaser}.
We use the following specification in this example:
\begin{minted}{js}
Dashboard {
    Sections: [{
        Metrics: ["Sales (SUM)", "Shipping Cost (SUM)"],
        DimensionGroups: [
            {PrimaryField: "Ship Date"}, 
            {PrimaryField: "Region"}
        ],
        MetricLayout: "Repeat"
    }]
}
\end{minted}
We have two metrics, Sales and Shipping Cost, along with two dimension groups. 
Each dimension group only has one dimension, Ship Date or Region.
Since we are using the \textit{Repeat} metric layout, each metric will be plotted in its own chart.
We combine every metric with every dimension group in a section so we end up with 4 charts total for this dashboard.
Part 3 of \autoref{fig:teaser} shows the generated dashboard.
We layout each metric on its own row; therefore we produce a grid of charts for this dashboard.
This grid layout allows users to compare metrics as well, such as comparing how Sales and Shipping Cost compare across Ship Date in the first column.

\subsubsection{Example 2: Comparing metrics in the same chart}

If we take the specification from \textit{Example 1} and change the \mintinline{js}{MetricLayout} value to \mintinline{js}{"Layer"} we can compare these two metrics on the same chart.
This is useful when trying to directly compare two metrics on a dashboard such as in \autoref{fig:example dashboards} (top) where we compare Sales and Shipping cost over Ship date all in the same chart.
In addition to changing the \mintinline{js}{MetricLayout}, we also add another dimension to the second dimension group.
Since we have two layered metrics across two dimension groups, we end up with two total charts in this dashboard.
The second chart in our dashboard has 4 columns total: the two metrics (layered on the same chart) of Sales and Shipping Cost along with the two dimensions of Ship Date and Region.
We can see how the simple abstraction of metrics and dimensions allows us to create sophisticated layerings of metrics and dimensions in a dashboard.

\subsubsection{Example 3: Putting it all together with multiple sections}
Finally, we can combine the previous two examples to create a rich dashboard with multiple sections.
Each section can have related or totally new metrics and dimensions.
The following specification creates the dashboard in \autoref{fig:example dashboards} (bottom) and demonstrates the flexibility of this specification to create rich dashboards just by combining metrics and dimensions into sections of dashboards.
\begin{minted}{js}
Dashboard {
    Sections: [{
        Metrics: ["Sales (SUM)", "Shipping Cost (SUM)"],
        DimensionGroups: [
            {PrimaryField: "Ship Date"}, 
            {PrimaryField: "Ship Date", SecondaryField: "Region"}
        ],
        MetricLayout: "Layer"
    },
    {
        Metrics: ["Sales (SUM)", "Shipping Cost (SUM)"],
        DimensionGroups: [
            {PrimaryField: "Region"}, 
            {PrimaryField: "Region", SecondaryField: "Category"}
        ],
        MetricLayout: "Repeat"
    }]
}
\end{minted}


\section{Discussion and Future Work}

In this section, we discuss the implications of \sys{} and the underlying specification and opportunities for future avenues of work.

\subsection{Focusing on data rather than visual specification}
Since the \sys{} UI allows users to focus on the aspects of their data they are interested in displaying, rather than \textit{how} to construct visualizations for this data, it speeds up the dashboard creation process.
Even if several of the automatically generated visualizations need to be tweaked, \sys{} can help users bootstrap their creation process.
We focus automation at the chart level, rather than the entire dashboard level. 
This is different than some other recommendation approaches that focus on recommending entire dashboards or multi-view visualizations \cite{Vartak2015SED, Wu2021MultiVisionDA}.
Users have the best idea of what combinations of data will be interesting to them but should not have to worry about how to actually encode those combinations.
By using a higher level of abstraction for dashboard creation, we view \sys{} as a first step towards more fluid and rapid dashboard authoring flows.
This makes it easier for a wider range of users to create dashboards, instead of just consuming them \cite{Tory2021FindingTD}.

\subsection{Interactive data-focused dashboard authoring}

When there exists an underlying specification for describing dashboards, it makes future authoring experiences easier to implement.
For example, in the current \sys{} system, the authoring flow is one way: users select their metrics and dimensions while viewing a preview of this generated dashboard, and then the dashboard is created and potentially tweaked.
Future authoring flows can support editing dashboards by selecting metrics and dimensions or creating dashboards in real-time from the specification, eliminating the need for a preview.
Regardless of the specification UI and flow, our dashboard specification offers a useful formalism for how to think about dashboards and their specification from data.



\subsection{Further incorporating best practices}
We can further incorporate design constraints for multi-view visualizations to make it easier for dashboard authors to follow best practices.
We obey constraints for axis consistency from \cite{Qu2018KeepingMV} since this makes interpretation easier, and plan on incorporating consistent color constraints in the future.
Our dashboard specification can be used independently of the chart recommender so improvements to chart recommendation will help make our generated dashboards better.
\section{Conclusion}
We present a data-first dashboard specification, where each section contains metrics combined with dimensions.
To support authoring dashboards with this specification, we created the \sys{} tool that provides a no-code user interface and chart recommendations.
We demonstrated the flexibility of the proposed specification and tool through examples of authoring three real-world dashboards.
Our specification of dashboards opens up new opportunities for making dashboard authoring experiences faster and easier.
\acknowledgments{
Thanks to the Visualization team at Databricks for their feedback on this work.
}

\bibliographystyle{abbrv-doi-hyperref}

\bibliography{refs}

\begin{thebibliography}{10}

\bibitem{googleLooker}
Google.
\newblock Google looker.
\newblock \url{ https://cloud.google.com/looker}, 2022.

\bibitem{Key2012VizDeckSD}
A.~Key, B.~Howe, D.~Perry, and C.~Aragon.
\newblock Vizdeck: Self-organizing dashboards for visual analytics.
\newblock In {\em Proceedings of the 2012 ACM SIGMOD International Conference
  on Management of Data}, SIGMOD '12, p. 681–684. Association for Computing
  Machinery, New York, NY, USA, 2012.
  \href{https://doi.org/10.1145/2213836.2213931}
{doi: {{%
10\hspace{.1pt}\discretionary{.}{%
}{.}\hspace{.4pt}1145\discretionary{/}{%
}{/}2213836\hspace{.1pt}\discretionary{.}{%
}{.}\hspace{.4pt}2213931}}}


\bibitem{maLADV2021}
R.~Ma, H.~Mei, H.~Guan, W.~Huang, F.~Zhang, C.~Xin, W.~Dai, X.~Wen, and
  W.~Chen.
\newblock Ladv: Deep learning assisted authoring of dashboard visualizations
  from images and sketches.
\newblock {\em IEEE Transactions on Visualization and Computer Graphics},
  27(9):3717--3732, 2021. \href{https://doi.org/10.1109/TVCG.2020.2980227}
{doi: {{%
10\hspace{.1pt}\discretionary{.}{%
}{.}\hspace{.4pt}1109\discretionary{/}{%
}{/}TVCG\hspace{.1pt}\discretionary{.}{%
}{.}\hspace{.4pt}2020\hspace{.1pt}\discretionary{.}{%
}{.}\hspace{.4pt}2980227}}}


\bibitem{mackinlayShowME2007}
J.~Mackinlay, P.~Hanrahan, and C.~Stolte.
\newblock Show me: Automatic presentation for visual analysis.
\newblock {\em IEEE Transactions on Visualization and Computer Graphics},
  13(6), 2007. \href{https://doi.org/10.1109/TVCG.2007.70594}
{doi: {{%
10\hspace{.1pt}\discretionary{.}{%
}{.}\hspace{.4pt}1109\discretionary{/}{%
}{/}TVCG\hspace{.1pt}\discretionary{.}{%
}{.}\hspace{.4pt}2007\hspace{.1pt}\discretionary{.}{%
}{.}\hspace{.4pt}70594}}}


\bibitem{powerBI}
Microsoft.
\newblock Microsoft powerbi.
\newblock \url{https://powerbi.microsoft.com/en-us/}, 2022.

\bibitem{Pandey2022MEDLEYIR}
A.~Pandey, A.~Srinivasan, and V.~Setlur.
\newblock Medley: Intent-based recommendations to support dashboard
  composition.
\newblock {\em IEEE Transactions on Visualization and Computer Graphics},
  29(1):1135--1145, 2023. \href{https://doi.org/10.1109/TVCG.2022.3209421}
{doi: {{%
10\hspace{.1pt}\discretionary{.}{%
}{.}\hspace{.4pt}1109\discretionary{/}{%
}{/}TVCG\hspace{.1pt}\discretionary{.}{%
}{.}\hspace{.4pt}2022\hspace{.1pt}\discretionary{.}{%
}{.}\hspace{.4pt}3209421}}}


\bibitem{Qu2018KeepingMV}
Z.~Qu and J.~Hullman.
\newblock Keeping multiple views consistent: Constraints, validations, and
  exceptions in visualization authoring.
\newblock {\em IEEE Transactions on Visualization and Computer Graphics},
  24(1):468--477, 2018. \href{https://doi.org/10.1109/TVCG.2017.2744198}
{doi: {{%
10\hspace{.1pt}\discretionary{.}{%
}{.}\hspace{.4pt}1109\discretionary{/}{%
}{/}TVCG\hspace{.1pt}\discretionary{.}{%
}{.}\hspace{.4pt}2017\hspace{.1pt}\discretionary{.}{%
}{.}\hspace{.4pt}2744198}}}


\bibitem{Roberts1998OnEM}
J.~C. Roberts.
\newblock On encouraging multiple views for visualization.
\newblock {\em Proceedings. 1998 IEEE Conference on Information Visualization.
  An International Conference on Computer Visualization and Graphics (Cat.
  No.98TB100246)}, pp. 8--14, 1998.

\bibitem{Sarikaya2019WhatDW}
A.~Sarikaya, M.~Correll, L.~Bartram, M.~Tory, and D.~Fisher.
\newblock What do we talk about when we talk about dashboards?
\newblock {\em IEEE Transactions on Visualization and Computer Graphics},
  25(1):682--692, 2019. \href{https://doi.org/10.1109/TVCG.2018.2864903}
{doi: {{%
10\hspace{.1pt}\discretionary{.}{%
}{.}\hspace{.4pt}1109\discretionary{/}{%
}{/}TVCG\hspace{.1pt}\discretionary{.}{%
}{.}\hspace{.4pt}2018\hspace{.1pt}\discretionary{.}{%
}{.}\hspace{.4pt}2864903}}}


\bibitem{srinivasan2023bolt}
A.~Srinivasan and V.~Setlur.
\newblock {BOLT: A Natural Language Interface for Dashboard Authoring}.
\newblock In T.~Hoellt, W.~Aigner, and B.~Wang, eds., {\em EuroVis 2023 - Short
  Papers}. The Eurographics Association, 2023.
  \href{https://doi.org/10.2312/evs.20231035}
{doi: {{%
10\hspace{.1pt}\discretionary{.}{%
}{.}\hspace{.4pt}2312\discretionary{/}{%
}{/}evs\hspace{.1pt}\discretionary{.}{%
}{.}\hspace{.4pt}20231035}}}


\bibitem{tableau}
Tableau.
\newblock Tableau dashboards.
\newblock \url{https://www.tableau.com/learn/get-started/dashboards}, 2022.

\bibitem{Tory2021FindingTD}
M.~Tory, L.~Bartram, B.~Fiore-Gartland, and A.~Crisan.
\newblock Finding their data voice: Practices and challenges of dashboard
  users.
\newblock {\em IEEE Computer Graphics and Applications}, 43(1):22--36, 2023.
  \href{https://doi.org/10.1109/MCG.2021.3136545}
{doi: {{%
10\hspace{.1pt}\discretionary{.}{%
}{.}\hspace{.4pt}1109\discretionary{/}{%
}{/}MCG\hspace{.1pt}\discretionary{.}{%
}{.}\hspace{.4pt}2021\hspace{.1pt}\discretionary{.}{%
}{.}\hspace{.4pt}3136545}}}


\bibitem{Tufte01}
E.~R. Tufte.
\newblock {\em The Visual Display of Quantitative Information}.
\newblock Graphics Press, Cheshire, CT, 2 ed., 2001.

\bibitem{Vartak2015SED}
M.~Vartak, S.~Rahman, S.~Madden, A.~Parameswaran, and N.~Polyzotis.
\newblock Seedb: Efficient data-driven visualization recommendations to support
  visual analytics.
\newblock {\em Proc. VLDB Endow.}, 8(13), sep 2015.
  \href{https://doi.org/10.14778/2831360.2831371}
{doi: {{%
10\hspace{.1pt}\discretionary{.}{%
}{.}\hspace{.4pt}14778\discretionary{/}{%
}{/}2831360\hspace{.1pt}\discretionary{.}{%
}{.}\hspace{.4pt}2831371}}}


\bibitem{Baldonado2000GuidelinesFU}
M.~Q. Wang~Baldonado, A.~Woodruff, and A.~Kuchinsky.
\newblock Guidelines for using multiple views in information visualization.
\newblock In {\em Proceedings of the Working Conference on Advanced Visual
  Interfaces}, AVI '00, p. 110–119. Association for Computing Machinery, New
  York, NY, USA, 2000. \href{https://doi.org/10.1145/345513.345271}
{doi: {{%
10\hspace{.1pt}\discretionary{.}{%
}{.}\hspace{.4pt}1145\discretionary{/}{%
}{/}345513\hspace{.1pt}\discretionary{.}{%
}{.}\hspace{.4pt}345271}}}


\bibitem{2016-compassql}
K.~Wongsuphasawat, D.~Moritz, A.~Anand, J.~Mackinlay, B.~Howe, and J.~Heer.
\newblock Towards a general-purpose query language for visualization
  recommendation.
\newblock In {\em ACM SIGMOD Human-in-the-Loop Data Analysis (HILDA)}, 2016.
  \href{https://doi.org/10.1145/2939502.2939506}
{doi: {{%
10\hspace{.1pt}\discretionary{.}{%
}{.}\hspace{.4pt}1145\discretionary{/}{%
}{/}2939502\hspace{.1pt}\discretionary{.}{%
}{.}\hspace{.4pt}2939506}}}


\bibitem{compassQL2016}
K.~Wongsuphasawat, D.~Moritz, A.~Anand, J.~Mackinlay, B.~Howe, and J.~Heer.
\newblock Towards a general-purpose query language for visualization
  recommendation.
\newblock In {\em Proceedings of the Workshop on Human-In-the-Loop Data
  Analytics}, HILDA '16. Association for Computing Machinery, New York, NY,
  USA, 2016. \href{https://doi.org/10.1145/2939502.2939506}
{doi: {{%
10\hspace{.1pt}\discretionary{.}{%
}{.}\hspace{.4pt}1145\discretionary{/}{%
}{/}2939502\hspace{.1pt}\discretionary{.}{%
}{.}\hspace{.4pt}2939506}}}


\bibitem{2016-voyager}
K.~Wongsuphasawat, D.~Moritz, A.~Anand, J.~Mackinlay, B.~Howe, and J.~Heer.
\newblock Voyager: Exploratory analysis via faceted browsing of visualization
  recommendations.
\newblock {\em IEEE Trans. Visualization \& Comp. Graphics (Proc. InfoVis)},
  2016. \href{https://doi.org/10.1109/TVCG.2015.2467191}
{doi: {{%
10\hspace{.1pt}\discretionary{.}{%
}{.}\hspace{.4pt}1109\discretionary{/}{%
}{/}TVCG\hspace{.1pt}\discretionary{.}{%
}{.}\hspace{.4pt}2015\hspace{.1pt}\discretionary{.}{%
}{.}\hspace{.4pt}2467191}}}


\bibitem{Wu2021MultiVisionDA}
A.~Wu, Y.~Wang, M.~Zhou, X.~He, H.~Zhang, H.~Qu, and D.~Zhang.
\newblock Multivision: Designing analytical dashboards with deep learning based
  recommendation.
\newblock {\em IEEE Transactions on Visualization and Computer Graphics},
  28(1):162--172, 2022. \href{https://doi.org/10.1109/TVCG.2021.3114826}
{doi: {{%
10\hspace{.1pt}\discretionary{.}{%
}{.}\hspace{.4pt}1109\discretionary{/}{%
}{/}TVCG\hspace{.1pt}\discretionary{.}{%
}{.}\hspace{.4pt}2021\hspace{.1pt}\discretionary{.}{%
}{.}\hspace{.4pt}3114826}}}


\end{thebibliography}
\end{document}